
\input harvmac.tex

\Title{PUPT-1305}
{\vbox{\centerline{Quantum Rings and Recursion Relations}
	\vskip2pt\centerline{in 2D Quantum Gravity }}}
\centerline{Shamit Kachru\footnote{$^\dagger$}
{Supported in part by an NSF Graduate Fellowship}}
\bigskip\centerline{Joseph Henry Laboratories}
\centerline{Jadwin Hall}
\centerline{Princeton University}\centerline{Princeton, NJ 08544 USA}

\vskip .3in
     I study tachyon condensate perturbations to the action of the two
dimensional string theory corresponding to the c=1 matrix model.  These
are shown to deform the action of the ground ring on the
tachyon modules, confirming a conjecture of Witten. The ground ring
structure is used to derive recursion relations
which relate (N+1) and N tachyon  bulk scattering amplitudes.
These recursion relations allow one to compute all bulk
amplitudes.
\Date{January 1992} 

\newsec{Introduction}

     Since the discovery of the double scaling limit of random matrix
models [1][2][3], the study of $D\le 2$ quantum gravity has advanced
rapidly.  The most interesting of the exactly soluble models is
undoubtedly the c=1 model.  The continuum analogue of this model is a
two dimensional string theory, which in addition to one field theoretic
degree of freedom (the massless tachyon) contains an infinite number
of  discrete states [4].  These states appear only at certain quantized
values of the momentum, and some of them have been interpreted as
remnants of the transverse string excitations in the two dimensional
theory [5].  The currents which can be formed from the discrete states
have also been shown to generate a $W_\infty$  symmetry [6],[7].

     In [6], Witten has demonstrated that the ghost number 0, conformal
dimension (0,0) operators form a ring (with product given by the
operator product expansion) whose properties are particularly important
for understanding the model.  In this paper, I will show that the ring
structure allows one to find recursion relations which determine the
bulk tachyon scattering amplitudes.  I will restrict myself to the
non-compact string theory.  Similar recursion relations were found in
[8] using the Ward identities of the $W_\infty$ symmetry.

     The action of the unperturbed two dimensional string theory under
study is given, in conformal gauge, by:
$$S = {1 \over {2\pi}} \int {\sqrt {\hat g}} ( {\hat g}^{ab}\partial_a
\phi\partial_b\phi + {\hat g}^{ab}\partial_aX\partial_bX - {Q\over 4}{\hat
R} \phi + 2\mu e^{-\sqrt2 \phi} ) \eqno (1)$$
Here, Q = $2\sqrt 2$ and $\mu$ is the cosmological constant, $\hat g$
is a fiducial metric and $\hat R$ is its curvature scalar, X is a free
boson, and $\phi$ is the Liouville field.

\newsec{Tachyon Modules of the Ground Ring}

     It was shown in [6] that in the non-compact model, the ground ring
is the ring generated by $a_+$ and $a_-$ where
$$\displaystyle a_+ = \vert \displaystyle ( cb + {i \over \sqrt 2}(\partial X -
i\partial\phi))\vert^{2}
\displaystyle e^{{1 \over \sqrt {2}}(iX - \phi)}\eqno (2)$$
$$\displaystyle a_- = \vert \displaystyle ( cb -{i \over \sqrt 2}(\partial X +
i\partial\phi)) \vert^{2}
\displaystyle e^{{-1 \over \sqrt {2}}(iX + \phi)} $$
It is further argued there that one should expect the ground ring
generators to satisfy a relation
$$ a_{+}a_{-} = \mu \eqno (3) $$
heuristically on the grounds that $a_+$, $a_-$ are analogous to p+q, p-q
in the fermion field theory avatar of the matrix model [9], where
(3) becomes the equation defining the fermi surface.  In addition, an
ansatz is given in [6] for how the relation (3) should be deformed when
the action (1) is perturbed by  discrete state conformal dimension
(1,1) moduli.

     It is shown in [10] that the tachyon states of the theory fall into
modules of the ring.  The tachyon states are given by:
$$T_{q}^{\pm} = e^{{1 \over \sqrt {2}}[iqX + (2 \mp q )\phi]}
 \eqno (4)
$$
and for vanishing $\mu$, the regular tachyon modules (the integer q
tachyons are  discrete states) are specified (for $+$ tachyons) by:
$$\displaystyle a_{+}c{\bar c} T_{q}^{+} = q^{2}c{\bar c} T_{q+1}^{+} + [Q,..]
\eqno (5)$$
$$\displaystyle a_{-}c{\bar c} T_{q}^{+} = 0 + [Q,..] $$
where Q is the BRST operator. The $-$ tachyon modules are similar, with the
roles of
the two ring generators being reversed.

     Notice in particular that the relation (3) is satisfied for
$\mu = 0$  on the tachyon modules.  We will see that its conjectured
generalizations will also be satisfied on the regular tachyon modules
(this is also discussed in [10]) and that the structure of the deformed modules
when the action (1) is
perturbed will allow us to derive recursion relations for the tachyon
correlation functions.

\newsec {Perturbations of S and the Deformed Ground Ring}

     Recall that in string theory, conformal dimension (1,1) operators
are moduli which can be integrated over the worldsheet, and hence may be
added to the Lagrangian.

     Let us begin by considering the addition to the action (1) of an
arbitrary $+$  tachyon
condensate:
$$ S' = S_{\mu = 0} + t_{k} \int T_k^+ \eqno (6)$$
How are the modules (5) deformed by this perturbation to the action?

     The lowest order correction to the action of $a_{+}a_{-}$ on the
module containing $T_{q}^{+}$ comes from the order $t_{k}$ correction
to the action of $a_{-}$.  Expanding the $e^{-S'}$ in the functional
integral representation for correlation functions, it is apparent that
this lowest order correction is:
$$ a_{-}c{\bar c}T_{q}^{+} = - lim_{w \rightarrow 0}c{\bar c}T_{q}^{+}(0)
a_{-}(w)  t_{k} \int T_{k}^{+}(z) d^{2}z\eqno (7)$$
Performing the contractions, we see that the leading term is:
$$-t_{k} \int d^{2}z |z|^{2q+2k-4} |z-1|^{-2k} c{\bar c} T_{q+k-1}^{+}(0)
=\eqno
(8)$$
$$-\pi t_{k}{\Gamma (q+k-1) \Gamma (1-q) \Gamma (1-k)\over {\Gamma (2-q-k)
\Gamma
(q) \Gamma (k)}} c{\bar c}T_{q+k-1}^{+}(0) $$
Defining $$\Delta (q) ={ \Gamma (q) \over \Gamma (1-q)}$$ we find:
$$a_{-}c{\bar c}T_{q}^{+} = - t_{k} \pi \Delta (k)^{-1} \Delta (q)^{-1} \Delta
(q+k-1) T_{q+k-1}^{+} \eqno (9)$$
This is the desired first order correction to the $a_{-}$ action on
the tachyon module.  However, from [11] [12] we expect the correlation
functions etc. of the two dimensional string theory to look simple only
in terms of the renormalized tachyon field and couplings, $\tilde
T_{q}^{+}$ and $\tilde t_{k}$:
$$\tilde T_{q}^{+} =  \Delta (q) T_{q}^{+}\eqno(10)$$
$$\tilde t_{k} = \pi  \Delta (k)^{-1} t_{k}$$
Interestingly, it has been determined in [13] that it is exactly with
this normalization that the tachyon transforms like a typical
dimension 1 field under a Virasoro sub-algebra of the algebra of the
$W_{\infty}$ charges.

     In terms of these renormalized fields and couplings, the deformed action
of the ground ring generators on the tachyons $T_{q}^{+}$ in the
theory with perturbed action (6) becomes:
$$a_{+} c{\bar c}\tilde T_{q}^{+} = - c{\bar c}\tilde T_{q+1}^{+} + [Q,..]
\eqno (11)$$
$$a_{-} c{\bar c} \tilde T_{q}^{+} = - \tilde t_{k} c{\bar c}\tilde
T_{q+k-1}^{+} + [Q,..]$$
The case of interest for the ansatz of [6] regarding deformations to the
fermi surface (3) is the case when k is positive integral, and the
deformation is by one of the discrete states.  In this case, the
multiplicative renormalization (10) becomes singular, but proceeding
with this caveat we find from (11) that:
$$a_{+}a_{-}c{\bar c}\tilde T_{q}^{+} = -a_{+} \tilde t_{k}c{\bar c}\tilde
T_{q+k-1}^{+} =
(-1)^{k}\tilde t_{k} (a_{+})^{k} \tilde T_{q}^{+} \eqno(12)$$
which implies that acting on the tachyon states c${\bar c}\tilde T_{q}^{+}$ the
ring generators satisfy the relation:
$$ a_{+}a_{-} = (-1)^{k}\tilde t_{k} (a_{+})^{k}\eqno(13)$$
In the special case $k=0$ and denoting $\tilde t_{0}$ by $\mu$ this
reproduces (3), confirming this part of the analogy of 2d string theory
with the matrix model.  Notice also that by considering perturbations
by $\tilde T_{k}^{-}$ , we would find a very similar
relation with the roles of $a_{-}$ and $a_{+}$ reversed.

\newsec {Recursion Relations for Tachyon Correlation Functions}

     Using the relations (7),(8), we see that in essence $a_{-}$ can be
used to fuse two $+$ tachyons into a single $+$
tachyon, while we know that for $-$ tachyons:
$$a_{-}T_{p}^{-} = p^{2} T_{p-1}^{-} + [Q,..]\eqno (14)$$
In addition, since $a_{-}$ is a conformal dimension (0,0) BRST invariant
operator, correlation functions with $a_{-}$ insertions should be
independent of the position of the $a_{-}$ insertion, up to
contributions coming from contact terms at the boundaries of moduli space
(similar arguments were used in evaluating correlation functions with ground
ring insertions in [10],[14]).  I will now use this fact to find recursion
relations which allow us to compute all of the  bulk tachyon
correlation functions in terms of the three point function,  in similar
spirit to [8] where Ward identities of the $W_{\infty}$ symmetry were
used to find recursion relations.

     Consider the correlation function
$$ B_{n,1}(q_{i},p)  =  <a_{-}(z)c{\bar c}\tilde T_{q_{1}}^{+}(0) c{\bar
c}T_{p}^{-}(1) c{\bar c}\tilde T_{q_{2}}^{+}(\infty) \prod_{i=3}^n \int
d^{2}z_{i}\tilde T_{q_{i}}^{+}
>\eqno (15) $$
with n $+$ tachyons, one $-$ tachyon and a ground ring insertion. Notice that
$T_{p}^{-}$ remains unrenormalized,
as the conservation laws tell us that it  has integral momentum and
the renormalization factor (10) would be infinite.

     Before proceeding, it may help to discuss briefly some geometric
subtleties associated with correlation functions like $B_{n,1}$, which
contain insertions of operators at non-standard values of the ghost
number.  In string theory, when one considers normal correlation
functions of physical states which have ghost number two, the
correlation
functions on a genus $g$ Riemann surface with $n$ operator insertions yield
integrals of top forms on $\bar {\cal {M}}_{g,n}$, the Deligne-Mumford
compactification of the moduli space of Riemann surfaces of genus $g$ with
$n$ punctures.
In the case at hand, however, we are considering correlation functions
of $n$ operators where $n-1$ have standard ghost number, and one has
ghost number zero.  Hence, we are left with an element of $H^{6g-6+2n-2}
(\bar {\cal  {M}}_{g,n})$ which must be integrated over a homology cycle of
codimension
two in order to yield a number.  One would hope that the assertions made below
can be
formulated in these geometrical terms by saying that for certain natural
distinct choices of this homology cycle, in the specific cases at
hand, one obtains the same number. I do not understand how this works in
any detail, however, and it may not be the correct interpretation of the
underlying geometry.

Now, let us study $B_{n,1}$ in the
limits as $z \rightarrow 0$ and $z \rightarrow 1$.  The BRST commutators
which arise in  moving the ground ring insertion do not contribute to
the correlation function [10], so the two limits of the correlation
function must be equal.  However, we
cannot naively use the relations (5).  As we have seen, there will be
contributions from  contact terms fusing two tachyons into one.  These
can be interpreted as contributions from the BRST commutators in (5).

     Taking the limit as $z \rightarrow 1$ and using (14), we find:
$$ B_{n,1}(q_{i},p) = p^{2} <c{\bar c} \tilde T_{q_{1}}^{+}(0) c{\bar
c}T_{p-1}^{-}(1) c{\bar c} \tilde T_{q_{2}}^{+} (\infty) \prod_{i=3}^n \int
d^{2}z_{i}\tilde T_{q_{i}}^{+}>\eqno (16)$$
where from [10] we know that any further BRST commutators on the right
hand side of (14) do not contribute.  The limit as $z \rightarrow 0$ is
slightly more subtle, but using (7), (8) we realize that we simply pick
up (n-2) contributions where the integrated tachyons are fused with
$\tilde T_{q_{1}}^{+}$.  Hence:
$$ B_{n,1}(q_{i},p) = \sum_{i=3}^n \pi <c{\bar c} \tilde
T_{q_{1}+q_{i}-1}^{+}(0)
c{\bar c} T_{p}^{-}(1) c{\bar c} \tilde T_{q_{2}}^{+}(\infty) \prod_{j=3,
j\not=
i}^n \int d^{2}z_{j} \tilde T_{q_{j}}^{+}> \eqno (17)$$
Thus, combining (16), (17) we see that:
$$  p^{2} <c{\bar c} \tilde T_{q_{1}}^{+}(0)c{\bar
c}T_{p-1}^{-}(1) c{\bar c} \tilde T_{q_{2}}^{+} (\infty) \prod_{i=3}^n \int
d^{2}z_{i} \tilde T_{q_{i}}^{+}>\eqno (18)$$
$$ =  \sum_{i=3}^n \pi <c{\bar c} \tilde
T_{q_{1}+q_{i}-1}^{+}(0)
c{\bar c} T_{p}^{-}(1) c{\bar c} \tilde T_{q_{2}}^{+}(\infty) \prod_{j=3,
j\not=
i}^n \int d^{2}z_{j} \tilde T_{q_{j}}^{+}>$$
Denoting tachyon correlation functions with n plus tachyons
and 1 minus tachyon as (n,1) correlation functions, we see
that we have a recursion relation expressing an (n,1) correlation
function as a sum of (n-1,1) correlation functions.  It is obvious how
similar arguments can be used to obtain recursion relations for (1,n)
correlation functions, and it was shown in [10] that ring insertion
arguments of this sort also suffice to prove that all (n,m) correlation
functions with $n,m \ge 2$ vanish.

     It is well known that in fact the non-vanishing  bulk correlation
functions are constant (they are evaluated explicitly in [12]), and
if we normalize the (2,1) function $A_{2,1} = 1$ (henceforth $A_{n,1}$ will
be used to denote the (n,1) tachyon correlation function) we can find
the rest of the non-vanishing correlation functions using (18).

     From the conditions of momentum and Liouville momentum
conservation, we find that in $B_{n,1}$ in (15)
$$ |p| = n-2 \eqno (19)$$
Hence, using (18) to find $A_{3,1}$ we see that
$$ A_{3,1} = \pi A_{2,1} = \pi \eqno (20) $$
which we know to be correct.  Using the fact that we know $A_{n-1,1}$ to
be a constant, we can determine using (18) and (19) that:
$$ A_{n,1} = {\pi \over {n-2}} A_{n-1,1} \eqno (21)$$
and using $A_{2,1} = 1$ we find
$$A_{n,1} = {\pi^{n-2} \over {(n-2)!}} \eqno (22)$$
which reproduces the results of direct computation [11],[12].

\newsec {Conclusion}

     We have seen in this paper that the ground ring structure suffices
to determine all of the non-vanishing bulk correlation functions of
the two dimensional string theory (1).  It will be very interesting to
see if the ring structure in other two dimensional string backgrounds
(e.g., the two dimensional black hole [15])  yields similar
simplifications in calculating correlation functions. It would also be
gratifying to obtain a deeper understanding of the geometrical meaning of
the relations derived here.

$$ \underline {Acknowledgements} $$  I would like to thank Jacques
Distler for several enlightening discussions, Igor
Klebanov for pointing out that I should look for recursion relations, and
Edward
Witten for suggesting the problem and for his continual guidance.
$$ \bf REFERENCES $$

1. D. Gross and A. Migdal, \it Phys. Rev. Lett. \bf 64 \rm (1990), 127.

2. M. Douglas and S. Shenker, \it Nucl. Phys. \bf B335 \rm (1990), 635.

3. E. Brezin and V. Kazakov, \it Phys. Lett. \bf B236 \rm (1990), 144.

4. B. Lian and G. Zuckerman, \it Phys. Lett. \bf B254 \rm (1991), 417.

5. A. Polyakov, \it Mod. Phys. Lett. \bf A6 \rm (1991), 635.

6. E. Witten, IAS preprint IASSNS-HEP-91/51.

7. I. Klebanov and A. Polyakov, Princeton preprint PUPT-1281.

8. I. Klebanov, Princeton preprint PUPT-1302.

9. S. Das, A. Dhar, G. Mandal, and S. Wadia, IAS preprint IASSNS-HEP-51.

10. D. Kutasov, E. Martinec, and N. Seiberg, Rutgers preprint RU-91-49.

11. D. Gross and I. Klebanov, \it Nucl. Phys. \bf B359 \rm (1991), 3.

12. P. Di Francesco and D. Kutasov, Princeton preprint PUPT-1276.

13. E. Witten and B. Zwiebach, IAS preprint IASSNS-HEP-91/100.

14. M. Li, UCSB preprint UCSBTH-91-47.

15. E. Witten, \it Phys. Rev. \bf D44 \rm (1991), 314.

\end